\documentclass[aip, pof, twocolumn, 10pt,showkeys, superscriptaddress,showpacs,floatfix]{revtex4-2}
\usepackage{multirow}
\usepackage{enumitem}
\usepackage{amsfonts} 
\usepackage{amsmath}
\usepackage{amssymb}
\usepackage{graphicx}
\usepackage{subfigure}
\usepackage{color}
\usepackage{enumerate}
\usepackage[]{natbib}
\usepackage{sidecap}
\usepackage{soul}
\usepackage{cancel}
\usepackage{ulem}
\usepackage{dcolumn}%
\usepackage{bm}%
\usepackage{array} 
\usepackage{booktabs} 
\usepackage{fontspec}
\newcommand*\xbar[1]{%
  \hbox{%
    \vbox{%
      \hrule height 0.5pt 
      \kern0.5ex
      \hbox{%
        \kern-0.1em
        \ensuremath{#1}%
        \kern-0.1em
      }%
    }%
  }%
} 
\begin{document}

\title[]{Influence of finite-temperature degeneracy and superthermal ions on dust-acoustic solitary structures}

\author{Rupak Dey}
\email{rupakdey456@gmail.com }
\affiliation{Department of Basic Sciences and Humanities, Camellia Institute of Engineering \& Technology, MAKAUT, Burdwan – 713403, West Bengal, India.}
\author{Gadadhar Banerjee }%
\email[Corresponding author: ]{gadadhar@burdwanrajcollege.ac.in, gban.iitkgp@gmail.com}
\affiliation{Department of Mathematics, Burdwan Raj College, University of Burdwan, Burdwan-713 104, West Bengal, India.}

\date{\today}

\begin{abstract}
We examine dust-acoustic (DA) solitary structures in an unmagnetized, collisionless dusty electron–positron–ion (e-p-i) plasma in which electrons and positrons are described by finite-temperature Fermi–Dirac statistics and ions obey a superthermal $\kappa$-distribution. A normalized fluid–Poisson model is formulated using polylogarithm-based expressions for the partially degenerate electrons-positrons, while cold negatively charged dust grains provide the inertial response. Linear dispersion analysis yields a modified DA phase speed. Nonlinear solitary structures are investigated using the Sagdeev pseudopotential method. The system is found to support only negative potential (rarefactive) DA solitary waves within a bounded subsonic Mach number interval. The critical Mach number $M_c$, consequently, the corresponding linear DA speed show an explicit dependence on the degeneracy parameters and the ion spectral index. The amplitude and width of the solitary structures are shown to be highly sensitive to the electron (positron) degeneracy strength, the ion-positron concentration ratios, the ion temperature ratio, and the superthermality of the ions. A small-amplitude approximation of the pseudopotential reduces the system to the Korteweg-de Vries limit, providing closed-form expressions for the soliton characteristics, in agreement with the Sagdeev predictions. The results clarify the combined roles of finite-temperature degeneracy and superthermal ions in shaping nonlinear DA dynamics in space and astrophysical dusty plasmas.  
\end{abstract}

\maketitle

\section{INTRODUCTION}\label{Intro}
Dusty \cite{alfven1982origin} or complex plasmas consist of ionized electron–positron–ion (e–p–i) media embedded with massive, micron- or submicron-sized charged dust grains, which may carry either negative or positive charge  \cite{mamun2002solitary, ali2023attributes}. Such dusty e-p-i plasmas occur naturally in a broad range of space and astrophysical systems, including the magnetospheres of Earth and Jupiter  \cite{goertz1989dusty}, planetary ring systems such as those of Saturn, Uranus, and Neptune  \cite{ali2023attributes, yaroshenko2007dust}, active galactic nuclei  \cite{pandey2023broad}, and cometary comae and tails  \cite{mendis2013dusty}. They are also frequently generated in laboratory and industrial settings, such as plasma processing devices, glow discharges, complex plasma crystals, semiconductor fabrication, and plasma-chemical reactors  \cite{bailung2018characteristics, thomas1994plasma, anirudh20232022, fridman2008plasma, krasheninnikov2011dust}. Following the theoretical prediction of dust–acoustic solitary waves (DASWs) by Rao et al.  \cite{rao1990dust} and their subsequent experimental confirmation by Barkan et al. \cite{barkan1995laboratory}, extensive research has been devoted to understanding the linear and nonlinear dynamics of dust–acoustic waves (DAWs). This includes detailed investigations of solitary waves, shocks, vortices, and various phase-space structures  \cite{banerjee2015pseudopotential, mamun2002solitary, bharuthram1992large, el2024dust, ali2023attributes, ali2006dust, esfandyari2012large, baluku2011ion, bharuthram1992large, baluku2010dust}. It is worth noting that the interplay between linear dispersion and nonlinearity can give rise to another type of nonlinear structure, known as lump waves, offering deeper insight into complex dispersive phenomena in high-dimensional nonlinear systems. For a (2+1)-dimensional fourth-order nonlinear wave equation with five categories of nonlinear terms, soliton solutions are obtained using the Hirota bilinear method, and rational and lump solutions are derived through suitable limiting processes \cite{cheng2025soliton}. The fourth-order nonlinear wave equation is reducible to spatially symmetric models. A generalized bilinear form of the governing equation is derived using extended fourth-order bilinear derivatives, providing a convenient framework for analytical treatment \cite{ma2025dispersion}. In a recent work \cite{ma2025Kaup}, where a Kaup–Newell type matrix eigenvalue problem with four potentials is used to construct a Liouville integrable hierarchy by using the zero-curvature formulation, and integrability is established through a hereditary recursion operator and bi-Hamiltonian structure. The underlying formulation is compatible with higher-dimensional generalizations, suggesting that the proposed method can be extended to multi-dimensional nonlinear integrable systems.

In various dusty-plasma contexts, a variety of non-Maxwellian distribution models have been extensively employed to analyze linear and nonlinear wave propagation in diverse plasma environments, offering improved descriptions of systems where superthermal or nonlocal effects are significant  \cite{banerjee2015pseudopotential, baluku2011ion}. Light particle species are modeled using Maxwell–Boltzmann distributions  \cite{bharuthram1992large, esfandyari2012large}, often with ion and electron temperatures assumed to be comparable. Recent studies \cite{vaidya2025computational,vaidya2025surface,vaidya2024electro} have demonstrated that nonlinear transport and collective dynamics in structured fluid systems are strongly influenced by non-classical effects, such as surface modulation, rheological complexity, and externally driven pumping mechanisms. The characteristics of the induced electric field in an incompressible Casson fluid flowing through a microfluidic system are evaluated in terms of the electric potential function,
while the number density of the particles is described using the Boltzmann distribution. However, spacecraft observations have revealed non-Maxwellian features in the velocity distributions of electrons, positrons, and ions in space plasmas  \cite{baluku2010dust}. Assuming kappa-distributed ions and Maxwellian electrons, Shahzad et al. \cite{shahzad2023nonlinear} investigated DASWs in magnetized dusty plasmas by using reductive perturbation theory.  Using the Sagdeev potential formalism, El-Taibani and Wadati  \cite{el2024dust} showed that variations in ion superthermality markedly reshape the Mach-number domain, demonstrating its central role in the formation of arbitrary-amplitude DA solitary waves in dusty plasmas with positively charged grains. Banerjee and Maitra  \cite{banerjee2015pseudopotential} extended the analysis to plasmas with $\kappa$-distributed electrons and ions, showing that the non-Maxwellian (superthermal) characteristics introduced by the $\kappa$-distributions substantially modify the Mach-number bounds and the resulting solitary-wave profiles. 

In extremely dense environments, such as cosmic regions, compact astrophysical objects including white dwarfs, and active galactic nuclei, the number density of degenerate particles typically reaches $\sim 10^{28} - 10^{34}$ $\text{cm}^{-3}$ \cite{dey2022ion}. Under such conditions, the mean interparticle spacing becomes much smaller than the electron (or positron) thermal de Broglie wavelength, causing the particles to obey Fermi–Dirac (FD) statistics.  Electrons and positrons that obey FD statistics can become partially degenerate in dense or high-energy plasma environments \cite{landau2013course}. Partial electron (positron) degeneracy is expected in such dense plasma environments where the Fermi temperature becomes comparable to the thermal temperature. For electrons (positrons), the Fermi temperature is given by $T_F \simeq \hbar^2 \left(3\pi^2 n\right)^{2/3} /2 m k_B$, accordingly, partial degeneracy ($T \sim T_F$, or $\tau = T_F/T \lesssim 1$) arises for number densities in the range
$n \sim 10^{28}-10^{31}\,\mathrm{cm^{-3}}$ at temperatures
$T \sim 10^{6}-10^{8}\,\mathrm{K}$ \cite{shukla2011colloquium,landau2013course}. Such conditions are known to occur in white dwarf envelopes, magnetized neutron star environments, and dense regions of accretion flows, where electrons and positrons are neither fully classical nor completely degenerate. These considerations justify the adoption of finite-temperature Fermi–Dirac statistics for electrons and positrons in the present model. Their thermodynamic behavior is jointly governed by the chemical potential and the temperature, reflecting an intermediate state between classical nondegeneracy and full quantum degeneracy  \cite{dey2022ion, haas2016nonlinear}. In such regimes, electrons and positrons exhibit partial degeneracy, characterized by temperatures below or above their respective Fermi temperatures.  Consequently, these species are neither fully nondegenerate nor completely degenerate. Meanwhile, ions often develop marked superthermal populations that are accurately modeled by the Kappa distribution, in which the spectral index and temperature serve as independent parameters. This formulation, originally introduced by Vasyliūnas \cite{vasyliunas1968survey} to interpret satellite measurements, has since become a standard framework for describing non-Maxwellian ion populations in space plasmas. A recent investigation of solitary structures in partially degenerate e–p–i plasmas with classical cold ions revealed that both positron concentration and degeneracy effects play a decisive role in reshaping the permissible Mach-number interval and in regulating the soliton amplitude \cite{dey2022ion}. However, the arbitrary-amplitude dynamics of DASWs in partially degenerate dusty e–p–i plasmas with superthermal ions remain unexplored. In this work, we address this gap by investigating arbitrary-amplitude DASWs in a dusty e–p–i plasma where electrons and positrons are partially degenerate and ions obey a kappa distribution. The negatively charged dust grains supply the inertial response. For partially degenerate electrons-positrons, the density and pressure are computed from the full FD distribution by integrating over all momentum states ($-\infty < v < \infty$), with no imposed upper energy cutoff. Using the Sagdeev pseudopotential method, we analyze the existence domains, parameter dependence, and structural properties of the resulting solitary waves.

The paper is organized as follows. Section~\ref{Basic Equation} introduces the physical model and governing normalized equations for the dusty plasma model. The linear dispersion relation is presented in Sec.~\ref{Linear Wave Mode}. Section~\ref{Sagdeev Potential} develops the analysis of arbitrary and small-amplitude solitary waves using the Sagdeev pseudopotential approach. Finally,  Sec.~\ref{Conclusion} concludes the paper with a summary of the main findings and a discussion of future scope.


\section{BASIC EQUATIONS} \label{Basic Equation}
 We consider unmagnetized collisionless dusty e-p-i plasmas consisting of partially degenerate electrons and positrons, nondegenerate inertialess $\kappa$ - distributed ions, and negatively charged cold dust particles to study the propagation of the arbitrary amplitude dust-acoustic solitary waves (DASWs). Such a plasma composition is relevant to various astrophysical environments, including the envelopes of white dwarfs  \cite{shapiro2024black}, magnetospheres of neutron stars  \cite{michel1991theory}, supernova remnants, and accretion disks  \cite{vasyliunas1968survey}, where partially degenerate electron–positron (e-p) populations coexist with superthermal ions and charged dust grains  \cite{livadiotis2009beyond,draine2003interstellar,wickramasinghe2000magnetism}.The present analysis is carried out within a local electrostatic framework, neglecting external forces such as gravity, which is justified when the solitary-wave scale length is much smaller than the gravitational scale height. The fundamental set of normalized equations governing the dynamics of DA waves comprises the dust continuity equation, the dust momentum balance equation, and the Poisson equation, given by,
 \begin{equation}
 \frac{\partial {n_d}}{\partial t }+\nabla \cdot \left(n_d {\bf u}_d\right)=0,\label{eq-du-con}
 \end{equation}
  \begin{equation}
\frac{\partial {{\bf u}_d}}{\partial t }+\left({\bf u}_d\cdot\nabla\right){\bf u}_d=\nabla{\phi}, \label{eq-du-mom}
 \end{equation}
 \begin{equation}
 \nabla^2 \phi={\alpha_e n_e}-{\alpha_p n_p}-\alpha_i n_i+n_d.\label{eq-poi}
 \end{equation} 
The normalized variables and parameters used in this context are described progressively in subsequent discussions. In this study, electrons and positrons are considered partially degenerate and are described by Fermi–Dirac (FD) statistical distribution  \cite{dey2022ion}. We assume that the thermodynamic temperature of each species, denoted by $T_j$ (where $j = e$ for electrons and $j = p$ for positrons), is slightly higher than the corresponding Fermi-temperature $T_{F_j} = E_{F_j} / k_B$, i.e., $T_j > T_{F_j}$.  $k_B$ is the Boltzmann constant and $E_{F_j}$ is the Fermi energy of the $j$-th species. The normalized expressions for the number densities of inertialess electrons ($n_e$) and positrons ($n_p$), in terms of electrostatic potential ($\phi$) and unperturbed chemical potential ($\mu_{j0}$), are obtained as  \cite{dey2022ion},
 \begin{equation} \label{eq-dene}
n_e=\frac{Li_{3/2} \left[-\exp\left(\phi+\mu_{e0}\right)\right]}{Li_{3/2} \left[-\exp\left(\mu_{e0}\right)\right]},
\end{equation}
and
\begin{equation} \label{eq-denp}
n_p=\frac{Li_{3/2} \left[-\exp\left(-\phi/\sigma_p+\mu_{p0}\right)\right]}{Li_{3/2} \left[-\exp\left(\mu_{p0}\right)\right]},
\end{equation}
where $Li_\nu(-z)$ is the poly-logarithm function with index $\nu$, defined by  \cite{lewin1981polylogarithms},
\begin{equation}
Li_\nu(-z)=-\frac{1}{\Gamma(\nu)}\int_{0}^{\infty}\frac{s^{\nu-1}}{1+z^{-1}e^{s} }ds, \; \nu>0,
\end{equation}
and for $\nu < 0$ we apply the differential property
\begin{equation}
Li_\nu(-z)= z\frac{\partial }{\partial z} Li_{\nu+1}(-z), \; 1+\nu>0.
\end{equation} 
The equilibrium chemical potential $\mu_{j0}$ is related to the unperturbed number density $n_{j0}$ by the relation,
\begin{equation}
-\frac{n_{j0}}{Li_{3/2}[-\exp(\mu _{j0})]}\left(\frac{ m}{2\pi k_B T_j}\right)^{3/2}=2\left(\frac{m}{2\pi\hbar}\right)^3 ,\label{eq-den-chem}
\end{equation}
in which  $\mu_{j0}$ is given by  \cite{shukla2011colloquium},
\begin{equation}
-Li_{3/2}\left[-\exp({\mu_{j0}})\right]=\frac{4}{3\sqrt{\pi}}\tau_j^{3/2}, \label{eq-tem-ratio}
\end{equation}
where  $\tau_j\equiv{T_{F_j}}/{T_j}$. In the nondegenerate limit ($\tau_j \ll 1$), the degeneracy parameter $\mu_{j0}$ approaches to $-\infty$, while in the fully degenerate limit ($\tau_j \gg 1$), $\mu_{j0}$ tends to $+\infty$  \cite{shukla2011colloquium}. However, the present model is formulated within the limit of finite-temperature degeneracy ($\tau_j<1$), wherein the plasma resides in a partially degenerate state such that quantum statistical effects contribute to, but do not govern, the overall dynamical behavior.In this regime, Eq.~\eqref{eq-tem-ratio} simplifies  \cite{dey2022ion} to
\begin{equation}
\mu_{j0}\approx\ln{ \left[\frac{4}{3\sqrt{\pi}}\tau_j^{3/2}\right]};~j=e,~p. \label{eq-muj0}
\end{equation}

The normalized number density of the superthermal $\kappa$ - distributed ions is given by  \cite{banerjee2015pseudopotential}, 
\begin{equation} \label{eq-deni}
n_i=\left( 1+ \frac{\phi/\sigma_i}{\kappa_i-3/2} \right)^{-\kappa_i+1/2}, 
\end{equation}
where the parameter $\kappa_i$ controls the deviation from the standard Maxwellian distribution. Typically, the value of $\kappa_i$ is taken to be greater than $3/2$, which is appropriate for modeling the superthermal tails observed in space and astrophysical plasmas. Lower values of $\kappa_i$ indicate a more pronounced high-energy tail. In the extreme limit of very low $\kappa_i$ ($\kappa_i \to 3/2$), the kinetic effects associated with resonant particles and higher-order velocity moments become increasingly important. The present fluid–Poisson model assumes that the dominant dynamics are governed by low-order moments of the ion distribution, which remain well defined for moderately superthermal plasmas (typically, $\kappa_i \geq 3$), a range commonly adopted in space plasma studies. In this regime, collective dust–acoustic modes are adequately described by a fluid treatment with an effective $\kappa_i$ - modified pressure response. For very low values of $\kappa_i$, however, strong nonlocal and kinetic effects are expected, and a fully kinetic (Vlasov–Poisson) approach would be more appropriate. On the otherhand, in the limit $\kappa \to \infty$, the distribution reduces to the Maxwellian form, corresponding to thermal equilibrium.

By incorporating the effects of the finite temperature degeneracy of electrons and positrons, the generalized DA speed is derived as a modified expression of the classical DA speed, influenced by degeneracy parameters given by  \cite{haas2016nonlinear},
\begin{equation} \label{eq-DA-speed}
c_d=   \sqrt{\frac{Z_d}{m_d}\left(\frac{dp_e}{dn_e}\right)_0}= \sqrt{\frac{\beta_e Z_dk_BT_e}{m_d}},
\end{equation}
where $m_d$ denotes the dust mass, $Z_d$ is the number of charged dust particles, and $\beta_j=Li_{3/2} \left[-\exp\left(\mu_{j0}\right)\right]/{Li_{1/2} \left[-\exp\left(\mu_{j0}\right)\right]}$ for $j=e,~p$. The electron’s fluid
pressure $p_e = p(n_e)$  is specified by a barotropic equation of
state, which is given in Ref.  \cite{haas2016nonlinear}. Consequently, the generalized Debye length modified to $\lambda_{D}~ (= {c_d}/{\omega_{pd}})$.
In the nondegenerate limit, $\exp(\mu_{j0}) \ll 1$ for which $Li_\nu\left[-\exp(\mu_{j0})\right] \approx -\exp(\mu_{j0})$ and $\beta_j\approx1$, the expressions reduce to  $c_d \approx\left({ Z_d k_BT_e}/{m_d}\right)^{1/2}$,   $\lambda_{D}\approx \left({k_B T_e}/ {m_d\omega_{pd}^2} \right)^{1/2}$, and  the isothermal pressure law $p_j=n_jk_BT_j$, i.e.,  the well-known classical results are retrieved. In the opposite limit $\exp(\mu_{j0}) \gg 1$ for the fully degeneracy case $Li_\nu\left[-\exp(\mu_{j0})\right] \approx -\mu_{j0}^{\nu}/\Gamma(\nu+1)$, $\mu_{j0}\approx \tau_j$ and $\beta_j=(2/3)\tau_j$, one obtains $c_d=(2Z_d k_BT_{Fe}/3m_d)^{1/2}$,  $\lambda_D=(2k_BT_{Fe}/3m_d \omega_{pd}^2)^{1/2}$ and the Fermi pressure law: $p_j=(2/5)n_{j0}E_{Fj}(n_j/n_{j0})^{5/3}$.  The present study deals with the intermediate regime characterized by $\tau_j < 1$, where the chemical potential $\mu_{j0}$ is governed by Eq.~\eqref{eq-muj0}. In this case, the generalized DA speed is modified by a multiplicative factor $\beta_e$, a dimensionless quantity that scales the classical DA speed and varies with the parameter $\tau_j$.
Here, the number densities $n_j$ (with $j = d, i, e, p$ corresponding to dust, ions, electrons, and positrons, respectively) are normalized by their unperturbed values $n_{j0}$. The velocity of the dust fluid $\mathbf{u}_d$ is normalized by the DA speed, $(Z_d k_B T_e/m_d)^{1/2}$ and
 the electrostatic potential $\phi$ is normalized by $k_BT_e/e$.  The time and space variables are normalized by the dust plasma period $\omega_{pd}^{-1}=\left({{4\pi Z_d^2 n_{d0}e^2}/{m_d}}\right)^{-{1/2}}$  and the Debye length, $\lambda_{D}~ = ({Z_d k_B T_e}/{m_d \omega_{pd}^2})^{1/2}$, respectively.  Furthermore, $\sigma_i ={T_i}/{T_e}$,  $\sigma_p ={T_p}/{T_e}$, $\delta_i=n_{i0}/n_{e0}$, $\delta_p=n_{p0}/n_{e0}$, $\alpha_i={n_{i0}}/Z_d n_{d0}=\delta_i/(\delta_i+\delta_p-1)$, $\alpha_e={n_{e0}}/Z_d n_{d0}=1/(\delta_i+\delta_p-1)$, $\alpha_p={n_{p0}}/Z_d n_{d0}=\delta_p/(\delta_i+\delta_p-1)$ are the dimensionless quantities. The temperature ratios $\tau_e$   and $\tau_p$ are related by  $\tau_p= \delta_p^{2/3 }\tau_e/\sigma_p$\cite{dey2022ion}. As a result, any variation in the electron degeneracy parameter 
$\tau_e$ automatically induces a corresponding change in positron degeneracy. Therefore, in the subsequent analysis, the effects of electron degeneracy implicitly include the influence of positron degeneracy in a self-consistent manner. 


\section{LINEAR WAVE MODE}\label{Linear Wave Mode}
We investigate the propagation of DASWs in a dusty e-p-i plasma under the assumption of small-amplitude perturbations, where nonlinear effects can be neglected. To examine the existence and characteristics of the dust-acoustic (DA) mode, we derive the corresponding dispersion relation by linearizing the system of Eqs.~\eqref{eq-du-con}–\eqref{eq-deni} about the equilibrium state. This is achieved by considering first-order perturbations of the physical quantities relative to their unperturbed values as follows, 
$n_d=1+n_{d1}$,$n_i=1+n_{i1}$, $n_e=1+n_{e1}$, $n_p=1+n_{p1}$, ${\bf {u}}_d= {\bf{0}}+{\bf{ u}}_{d1} $ and $\phi=0+\phi_1$. A first-order approximation of the polylogarithm function is used  as,
\begin{equation}
\begin{split}
Li_\nu \left[-\exp[{\gamma(y_0+y_1)]}\right]&=Li_\nu\left[-\exp(\gamma y_0)\right]\\
 &+\gamma y_1 Li_{\nu-1}\left[-\exp(\gamma y_0)\right].
 \end{split}
\end{equation} 
Assuming the perturbations vary as plane waves of the form $\exp(i \mathbf{k} \cdot \mathbf{r} - i \omega t)$, where $\omega$ is the angular frequency normalized by the dust plasma frequency $\omega_{pd}$ and $\mathbf{k}$ is the wave vector normalized by the inverse of the Debye length $\lambda_D^{-1}$, we derive the dispersion relation as follows,
\begin{equation} \label{eq-disp}
\omega^2=\frac{k^2}{k^2+(\alpha_e/\beta_e)+\left(\alpha_p/\beta_p \sigma_p \right)+(\alpha_i C_i /\sigma_i)},
\end{equation}
where $C_i=(\kappa_i-1/2)/(\kappa_i-3/2)$. It is worth mentioning that, in the limiting case where the finite temperature degeneracy of electrons and positrons is neglected, the obtained dispersion relation \eqref{eq-disp} coincides with the result derived by Saberian et al. \cite{saberian2017nonlinear} for a dusty e–p–i plasma model incorporating superthermal electrons, positrons, and ions. Furthermore, in the limit $\kappa_i \to \infty$, which implies $C_i \to 1$, the dispersion relation \eqref{eq-disp} simplifies to,
\begin{equation} \label{eq-disp1}
\omega^2=\frac{k^2}{k^2+(\alpha_e/\beta_e)+\left(\alpha_p/\beta_p \sigma_p \right)+(\alpha_i /\sigma_i)},
\end{equation}
which corresponds to the case of Maxwellian-distributed ions. In the nondegenerate limit (i.e., when $\beta_j \approx 1$), the dispersion relation given by Eq.~\eqref{eq-disp1} further reduces to,
\begin{equation} \label{eq-disp2}
\omega^2 = \frac{k^2}{k^2 + \alpha_e + \alpha_p/\sigma_p + \alpha_i/\sigma_i},
\end{equation}
which represents the dispersion relation for DA waves in a dusty e–p–i plasma, assuming a Boltzmann-distributed electron, positron, and ion population  \cite{esfandyari2012large}.

\begin{figure}[!htbp] 
\centering
\includegraphics[width=3.4in,height=2.5in]{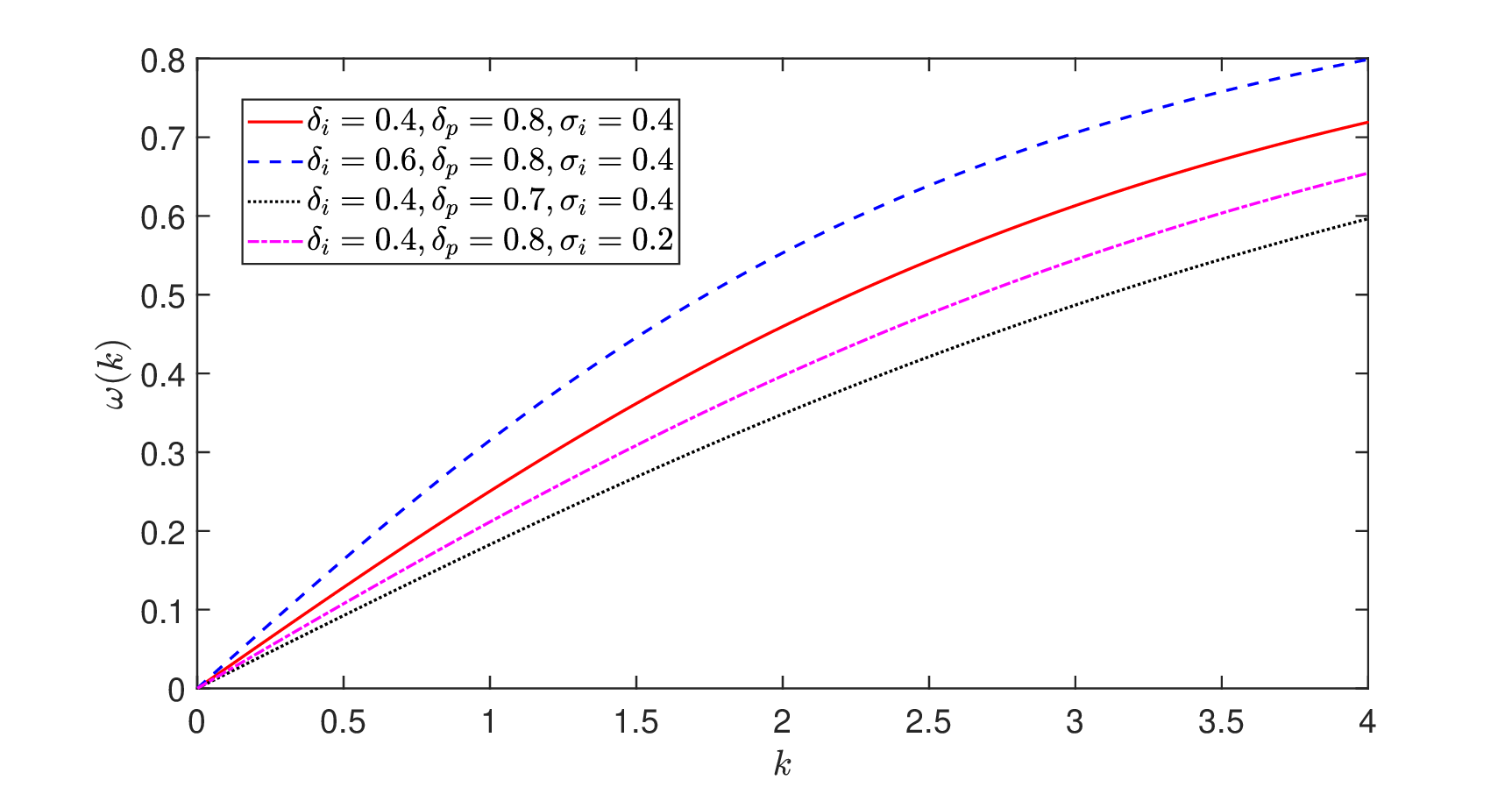}
\caption{ Plots of dispersion curves showing the variation of wave frequency $\omega$ with the wave number $k$ for different values of $\delta_i$, $\delta_p$ and $\sigma_i$. Here, $\tau_e=0.5$, $\kappa_i=5$ and $\sigma_p=1$.}
\label{fig:dispersion_1}
\end{figure}
 \begin{figure}[!htbp] 
\centering
\includegraphics[width=3.4in,height=2.5in]{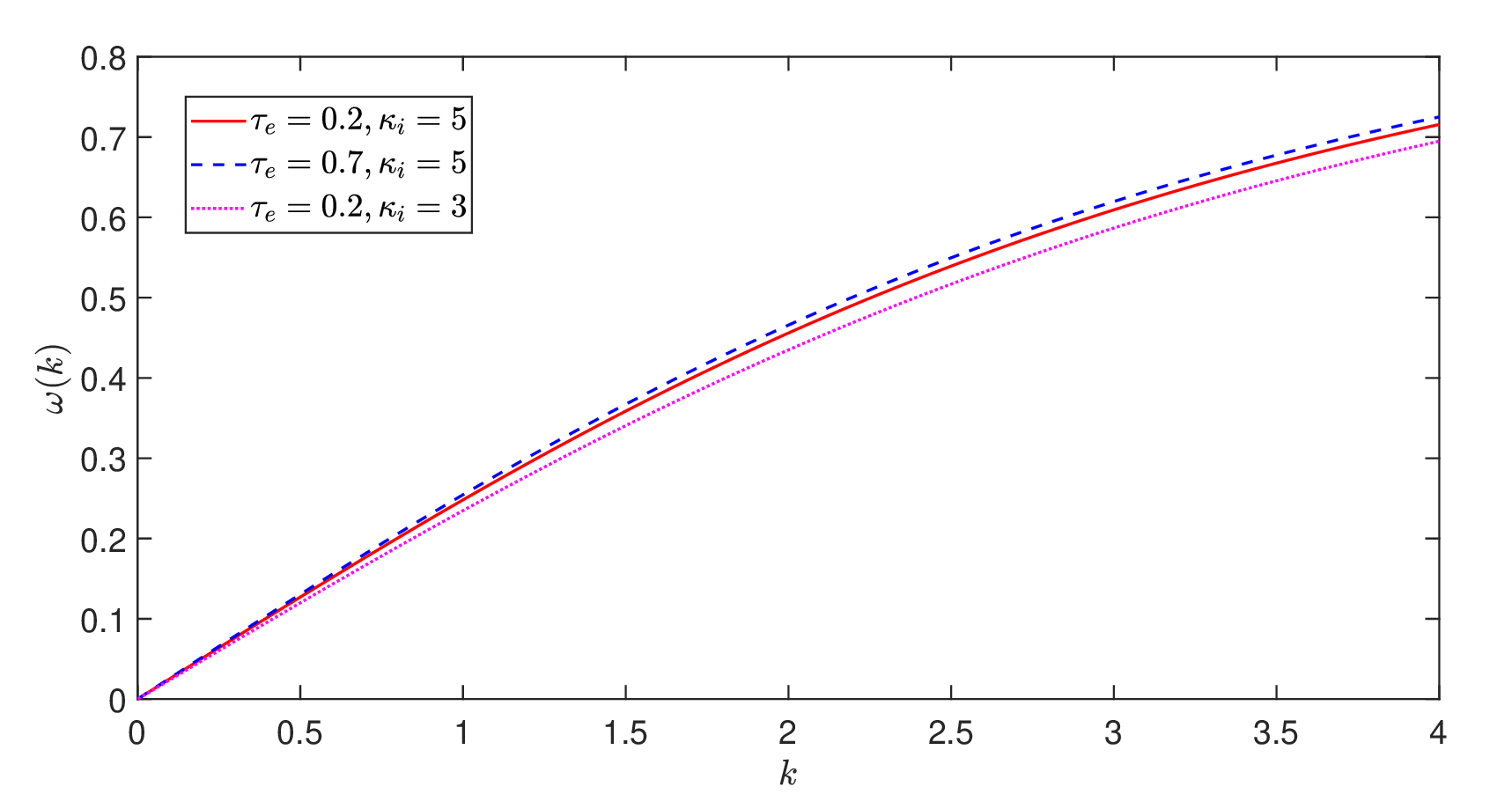}
\caption{ Plots of dispersion curves showing the variation of wave frequency $\omega$ with the wave number $k$ for different values of $\tau_e$ and $\kappa_i$. Here, $\delta_i=0.4$, $\delta_p=0.8$, $\sigma_i=0.4$ and $\sigma_p=1$.}
\label{fig:dispersion_2}
\end{figure}

The dispersion characteristics (dispersion relation (\ref{eq-disp})) of the DA waves in the dusty e-p-i plasma are illustrated in the Fig. \ref{fig:dispersion_1}, where the wave frequency $\omega$ is plotted as a function of the wave number $k$ for different values of plasma parameters $\delta_i$, $\delta_p$ and $\sigma_i$. Each of these parameters plays a significant role in modifying the collective wave behavior due to their impact on charge balance, pressure gradients, and overall plasma dynamics. An increase in $\delta_i$ leads to a noticeable increase in phase velocity across the entire wave number range.  As ions contribute to the inertia of the plasma, increasing $\delta_i$ enhances the ion response to the electric field, resulting in steeper dispersion curves. As $\delta_p$ increases, the wave frequency exhibits an increasing trend, which indicates that the inclusion of more positrons, which tend to neutralize electron effects, reduces the net charge imbalance and weakens the electrostatic restoring force. With increasing $\sigma_i$, the dispersion curve becomes steeper and more nonlinear, and it enhances the thermal pressure of the ions, contributing to stronger pressure gradients. A higher $\sigma_i$ effectively increases the acoustic speed of the ion, thus shifting the dispersion relation upward. 
Fig. \ref{fig:dispersion_2} illustrates the dispersion curves for different values of the electron degeneracy parameter $\tau_e$ and the ion spectral index $\kappa_i$. It is evident that the wave frequency $\omega$ increases monotonically with the wave number $k$ in all cases, indicating a normal dispersion behavior. It is observed that increasing the electron degeneracy parameter $\tau_e$ leads to an upward shift of the dispersion curve. This is expected because degenerate electrons (i.e. electrons with high Fermi energy compared to thermal energy) behave more collectively, contributing significantly to the dispersion properties of the wave. On the other hand, a reduction in the ion spectral index $\kappa_i$ results in a downward shift of the dispersion curve, signifying a decrease in the wave frequency. These parametric sensitivities highlight the competing roles of thermal electrons and superthermal ions in determining the propagation characteristics of dust–acoustic modes in multi-component dusty plasmas.

\section{NONLINEAR ANALYSIS}\label{Sagdeev Potential}
In the nonlinear regime, to investigate arbitrary amplitude DASWs, we employ the Sagdeev pseudopotential approach.   In this framework, all physical variables are considered as functions of a single normalized coordinate, $\zeta$, defined as $\zeta=l_xx+l_yy+l_zz-Mt$,   where $l_x$, $l_y$, and $l_z$ are the direction cosines of a line parallel to the coordinate axes. $M$ represents the Mach number, which quantifies the speed of the solitary wave relative to the DA speed $c_d$. Integrating Eqs. \eqref{eq-du-con} and \eqref{eq-du-mom}, and imposing  the appropriate boundary conditions  $u_d $, $\phi \to 0$, $n_d \to 1$ as $\zeta \to \pm\infty$, yields the expression for the dust number density,
\begin{equation}\label{eq-dendu}
n_d=\frac{M}{\sqrt{M^2+2\phi}}. 
\end{equation}
Substituting the expressions of the number densities of electrons, positrons, ions and dust particles from Eqs. \eqref{eq-dene}, \eqref{eq-denp}, \eqref{eq-deni} and \eqref{eq-dendu}, respectively, in Eq. \eqref{eq-poi}, we obtain,
\begin{equation}
    \begin{aligned}
   &\frac{d^2\phi}{d\zeta^2}=\alpha_e \frac{Li_{3/2} \left[-\exp\left(\phi+\mu_{e0}\right)\right]}{Li_{3/2} \left[-\exp\left(\mu_{e0}\right)\right]} \\
   & \left.-\alpha_p \frac{Li_{3/2} \left[-\exp\left(-\phi/\sigma_p+\mu_{p0}\right)\right]}{Li_{3/2} \left[-\exp\left(\mu_{p0}\right)\right]} \right. \\
  &  -\alpha_i \left( 1+ \frac{\phi/\sigma_i}{\kappa_i-3/2} \right)^{-\kappa_i+1/2} +\frac{M}{\sqrt{M^2+{2\phi}}} .
    \end{aligned}  \label{eq-energy}
    \end{equation}
Integrating Eq. \eqref{eq-energy} once and applying the boundary conditions $\phi \to 0$, $ {d\phi}/{d\zeta} \to 0$, and $ {d^2\phi}/{d\zeta^2}\to 0$ as $\zeta \to \pm\infty$, we obtain the energy integral describing a pseudoparticle of unit mass with velocity ${d\phi}/{d\zeta}$ and position $\phi$ as,  
    \begin{equation}
    \frac{1}{2}\left(\frac{d\phi}{d\zeta}\right)^2+V(\phi)=0,  \label{eq-energy-integral}
    \end{equation}
 where the pseudopotential $V(\phi)$ is given by,
    \begin{equation}
    \begin{aligned}
    V(\phi)= & \frac{\alpha_e}{Li_{3/2}\left[-\exp(\mu_{e0})   \right]}\\
    &\times \left\lbrace Li_{5/2}\left[-\exp{(\mu_{e0})}\right]- Li_{5/2}\left[-\exp{(\phi +\mu_{e0})}\right] \right\rbrace \\ 
&    +\frac{ \sigma_p \alpha_p}{Li_{3/2}\left[-\exp(\mu_{p0})   \right]} 
  \left\lbrace Li_{5/2}\left[-\exp(\mu_{p0})\right] \right.\\
  &\left. - Li_{5/2}\left[-\exp\left(-\phi/\sigma_p +\mu_{p0}\right)\right]\right\rbrace  \\
& + \alpha_i \sigma_i \left\lbrace 1-\left( 1+\frac{\phi/\sigma_i }{\kappa_i-3/2} \right)^{-\kappa_i+3/2} \right\rbrace \\ 
 & + M^2 \left\lbrace 1- \sqrt{1+\frac{2\phi}{M^2}} \right\rbrace.    
    \end{aligned} \label{eq-Sagdeev-Potential}
    \end{equation}
In what follows,  we outline the necessary conditions for the existence of arbitrary-amplitude DASWs, as discussed in  \cite{dey2022ion}. First, the Sagdeev pseudopotential $V(\phi)$ and its first derivative must vanish at the origin, i.e., $V(\phi)=0$ and $V'(\phi)=0$ at $\phi=0$, ensuring that the origin is an equilibrium point. Second, the second derivative of the potential must be negative at the origin, i.e. $ V''( \phi)<0$ at $\phi=0$, so that the equilibrium is unstable and a localized structure can emerge. Finally, there must exist a nonzero potential $\phi=\phi_m (\neq 0)$ such that $V(\phi)<0$ for $0<\lvert \phi \rvert<\lvert \phi_m \rvert$ and $V(\phi_m)=0$. This condition ensures that the pseudoparticle can oscillate within a potential well, with $\phi_m$ representing the maximum amplitude of the solitary wave or double layer, if such structures exist.


\subsection{ARBITRARY AMPLITUDE WAVE}\label{sec-arbitrary}
We verify the necessary conditions and investigate the existence of arbitrary amplitude DASWs in different parameter regimes. The first condition is trivially satisfied by Eq. (\ref{eq-Sagdeev-Potential}). The second criterion, $ V''( \phi)<0$ at $\phi=0$, is met provided that the Mach number $M$ exceeds a critical value $M_c$, given by
     \begin{equation}\label{eq-lower-mach}
 M_c=\left(\frac{\alpha_e}{\beta_e}+\frac{\alpha_p}{\beta_p \sigma_p}+\frac{\alpha_i C_i}{\sigma_i}\right)^{-1/2},
 \end{equation}
is referred to as the critical Mach number. This represents the lower bound on $M$ below which no solitary wave or double-layer solution can exist. The expression obtained for $M_c$ coincides with the phase velocity of DA waves derived from the linear dispersion relation (Eq.~\eqref{eq-disp}) in the long-wavelength limit, i.e. $k\ll 1$. This critical Mach number therefore corresponds to the phase velocity of linear DA waves as desired. In the non-degenerate limit, Eq. \eqref{eq-lower-mach} reduces to 
 \begin{equation}\label{eq-lower critical mach_1}
 M_c=\left(\alpha_e+\alpha_p / \sigma_p+\alpha_i C_i /\sigma_i \right)^{-1/2},
 \end{equation}
which can be further reduced to the case of Maxwellian-distributed
ions (i.e, when $\kappa_i \to \infty$) as follows,
\begin{equation}\label{eq-lower critical mach_2}
 M_c=\left(\alpha_e+\alpha_p / \sigma_p+\alpha_i /\sigma_i \right)^{-1/2}.
 \end{equation}
The expression \eqref{eq-lower critical mach_2} turns out to be identical with the result obtained in Ref.  \cite{saberian2017nonlinear} if both are expressed under the same normalisation.
Since the pseudopotential function $V(\phi)$ is real-valued 
and the poly-logarithm function $Li_{\nu}[z]$ converges for $|z|<1$, 
the function $V(\phi)$ exists in the interval $\phi_c^- \leq \phi \leq \phi_c^+$, 
where $\phi_c^-=max\{- M^2/2, \sigma_p\mu_{p0}, -(k_i-3/2)\sigma_i\}$  and $\phi_c^+=\mu_{e0}$.
We now incorporate the third criterion for the existence of SWs and double layers, which requires that there exist some $\phi_m (\neq 0)$ such that $V(\phi_m)=0$ and $V(\phi)<0$ for $0<|\phi|<|\phi_m|$. 
For such $\phi_m$ to exist, it is necessary that $V(\phi)$ possesses a local minima within the interval $0<|\phi|<|\phi_m|$. From the numerical investigation, it is found that $V(\phi)$ does not admit any local minima in the region $0<\phi<\phi_c^+$. However, in the range $\phi_c^-<\phi<0$,  local minima of $V(\phi)$  occur in various parametric regions. Consequently, negative potential solitary waves exist for certain typical values of the plasma parameters. For the existence of double layers, in addition to the solitary wave conditions, it is required that $V'(\phi_m)=0$ and $V''(\phi_m)<0$, i.e., $V(\phi)$ have a local maximum at $\phi=\phi_m$. Numerical analysis shows that $V(\phi)$ remains positive for $\phi_c^-<\phi<\phi_m$, and therefore double layers do not exist in such partially degenerate dusty e-p-i plasmas. The critical Mach number $M_c$ represents the lower bound of the Mach number for the existence of negative potential SWs. The upper bound can be obtained from the relation $V(\phi_c^{-})=0$. However, it is not easy to express this upper limit $M_u$ explicitly and must be determined numerically in terms of the relevant plasma parameters. In our analysis, we numerically obtained the upper limit of Mach number for the existence of negative potential SWs. The admissible domains of the Mach number, bounded by the critical value $M_c$ and the upper limit $M_u$ for the existence of a negative potential DASW, are presented in Table \ref{table_Mach} for different values of plasma parameters. By carefully examining the admissible regions of the Mach number, we observe that DA solitary structures are restricted to values of $M<1$. This clearly indicates that the system supports only subsonic DA solitary waves, while the formation of supersonic solitary structures is prohibited within the considered plasma configuration. Such a restriction arises because of the combined influence of the plasma parameters $\kappa_i$ and $\tau_e$, which prevents the existence of solitary solutions beyond the subsonic regime.

\begin{figure}[!htbp] 
\centering
\includegraphics[width=3.2in,height=2.5in]{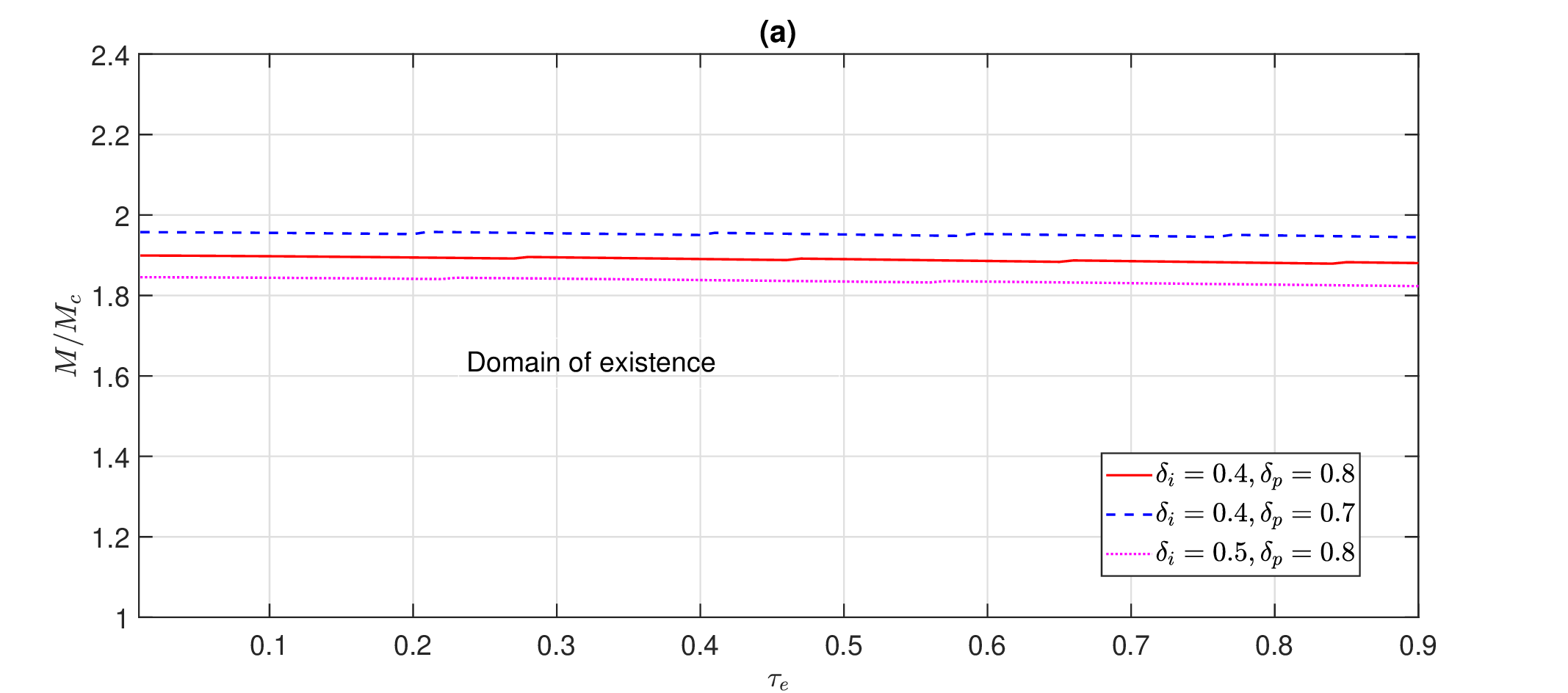}
\includegraphics[width=3.2in,height=2.5in]{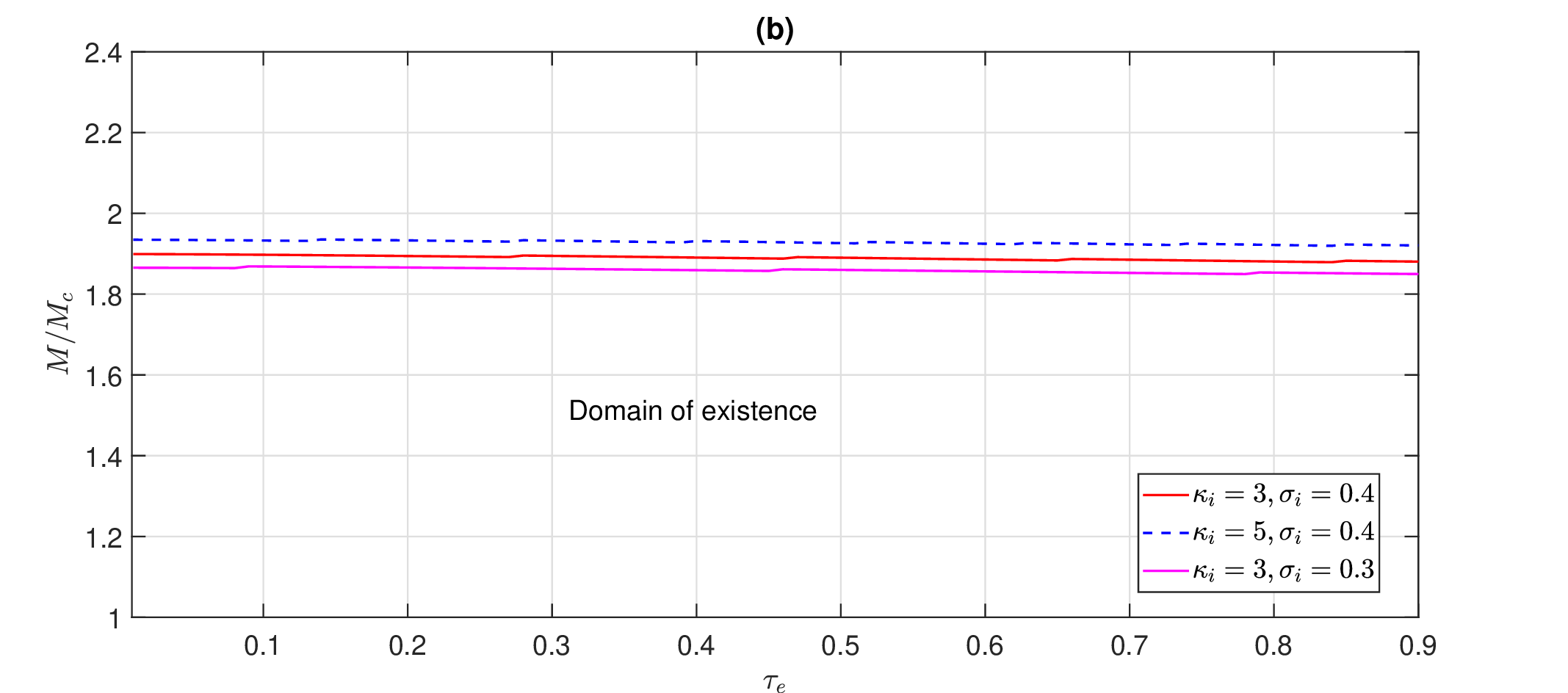}
\caption{Domain of existence of negative potential solitary waves in terms of the true Mach number $M/M_c$ and $\tau_e$ for different values of (a) $\delta_i$, $\delta_p$ where $\kappa_i=3$, $\sigma_i=0.4$ and  (b) $\kappa_i$, $\sigma_i$ where $\delta_i=0.4$, $\delta_p=0.8$. Here, $\sigma_p=1$.}
\label{fig:mach_number}
\end{figure}

The Mach number $M$ is normalized with respect to the DA speed $c_d$, and the system parameters govern the variation of the critical Mach number $M_c$. It should be emphasized that, since $M$ is defined through a specific normalization, caution is required in its physical interpretation. To avoid this ambiguity, the true Mach number is defined as the ratio $M/M_c$, by which the reference speed $c_d$ is eliminated  \cite{banerjee2016arbitrary}. Thus, the existence criterion for solitary waves $M>M_c$ reduces to $M/M_c>1$. Consequently, the admissible domain of the true Mach number is given by $1<M/M_c<M_u/M_c$. In particular, corresponding to the parametric values listed in the first row of Table \ref{table_Mach} the true Mach number lies within the range $1 \leq M/M_c \lesssim 1.894$. Henceforth, the true Mach number has been employed for plotting various figures. The domain of existence of negative potential solitary waves is presented in Fig.  \ref{fig:mach_number} in terms of the true Mach number,  $M/M_c$ and the temperature ratio $\tau_e$ for different values of the parameters as mentioned in the legend. In Fig. \ref{fig:mach_number}(a), the effect of the density ratios $\delta_i$ and $\delta_p$ is depicted for fixed values of $\kappa_i=3$ and $\sigma_i=0.4$. On the other hand, in Fig. \ref{fig:mach_number}(b) the effect of $\kappa_i$ and $\sigma_i$ is depicted for fixed values of $\delta_i=0.4$ and $\delta_p=0.8$. The profiles of $M_u/M_c$ indicate the weak dependence of $\tau_e$. However, the plots exhibit a few minor jumps at certain values of $\tau_e$, causing a slight shift in the admissible domain of Mach numbers. It is observed that increasing both density ratios leads to a narrowing of the existence domain. Fig. \ref{fig:mach_number}(b) reveals that larger values of $\kappa_i$  and $\sigma_i$ enhance the existence domain of solitary waves. Thus, enhanced superthermal effects (lower $\kappa_i$) and cooler ions (smaller $\sigma_i$) broaden the accessibility of negative potential SWs.

\begin{table}[!htbp]
\centering
\setlength{\tabcolsep}{10pt} 
\renewcommand{\arraystretch}{1.2} 
\begin{tabular}{l l l l l c}
\hline
$\delta_i$ & $\delta_p$ & $\sigma_i$ & $\kappa_i$ & $\tau_e$ & $M$ \\
\hline
0.4 & 0.8 & 0.4 & 3 & 0.2 & 0.242--0.458 \\ 
0.4 & 0.8 & 0.4 & 3 & 0.6 & 0.246--0.464 \\ 
0.4 & 0.8 & 0.4 & 5 & 0.2 & 0.256--0.596 \\ 
0.4 & 0.8 & 0.3 & 3 & 0.2 & 0.224--0.418 \\ 
0.4 & 0.7 & 0.4 & 3 & 0.2 & 0.173--0.339 \\
0.5 & 0.8 & 0.4 & 3 & 0.2 & 0.279--0.515 \\
\hline
\end{tabular}
\caption{The admissible domains of Mach number for the existence of negative potential DASW are presented for different parameter values. Here $\sigma_p=1$.}
\label{table_Mach}
\end{table}

\begin{figure}[!htbp] 
\centering
\includegraphics[width=3.4in,height=2.5in]{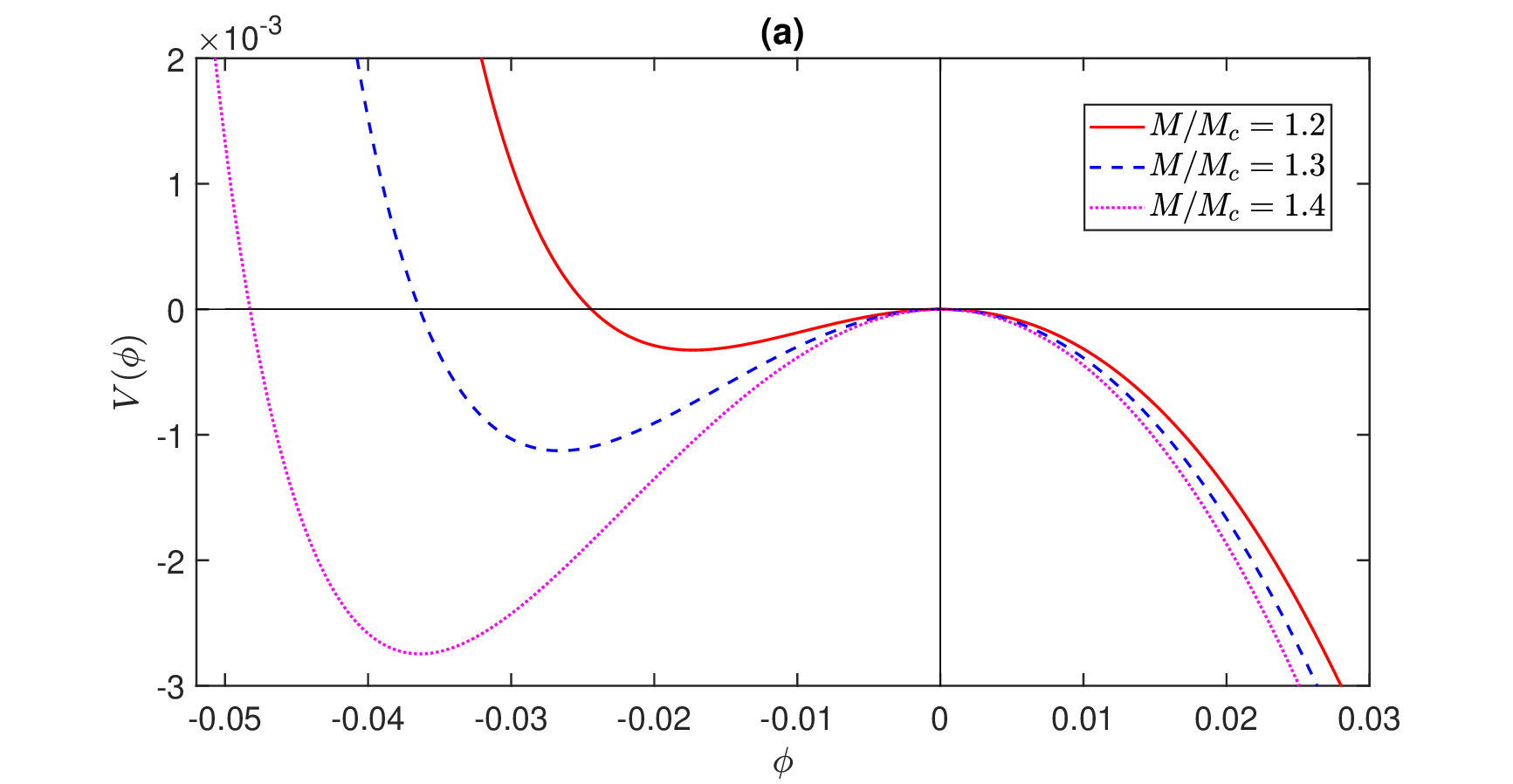}
\includegraphics[width=3.4in,height=2.5in]{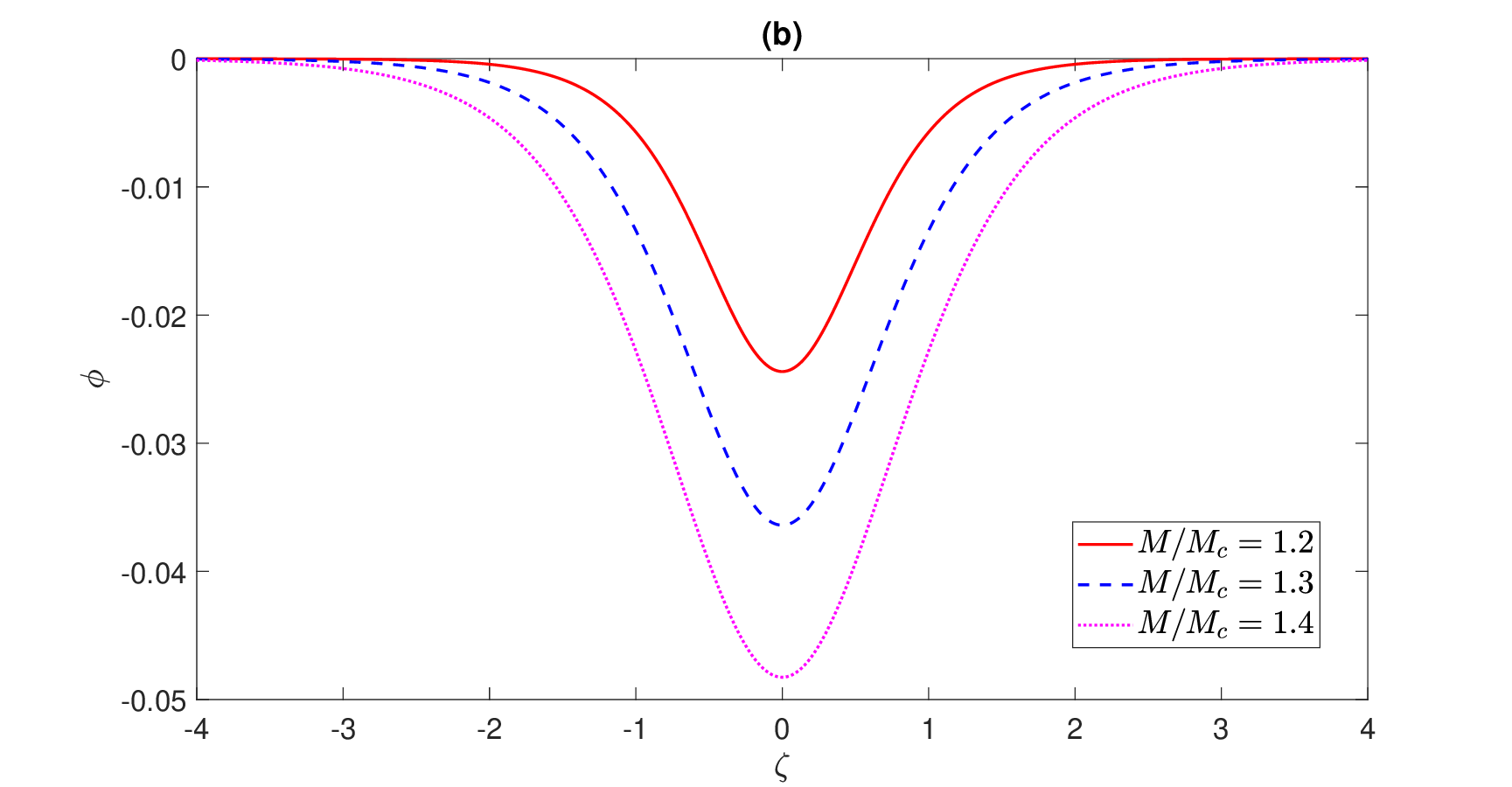}
\caption{(a) The Sagdeev pseudopotential $V(\phi)$ and (b) the corresponding solitary-wave potential profiles $\phi$ are shown for different values of the true Mach number $M/M_c$. The parameters used are $\kappa_i=3$, $\sigma_i=0.4$, $\delta_i=0.4$, $\delta_p=0.8$, $\sigma_p=1$ and $\tau_e=0.2$.}
\label{fig:Vphi_M}
\end{figure}
\begin{figure}[!htbp] 
\centering
\includegraphics[width=3.4in,height=2.5in]{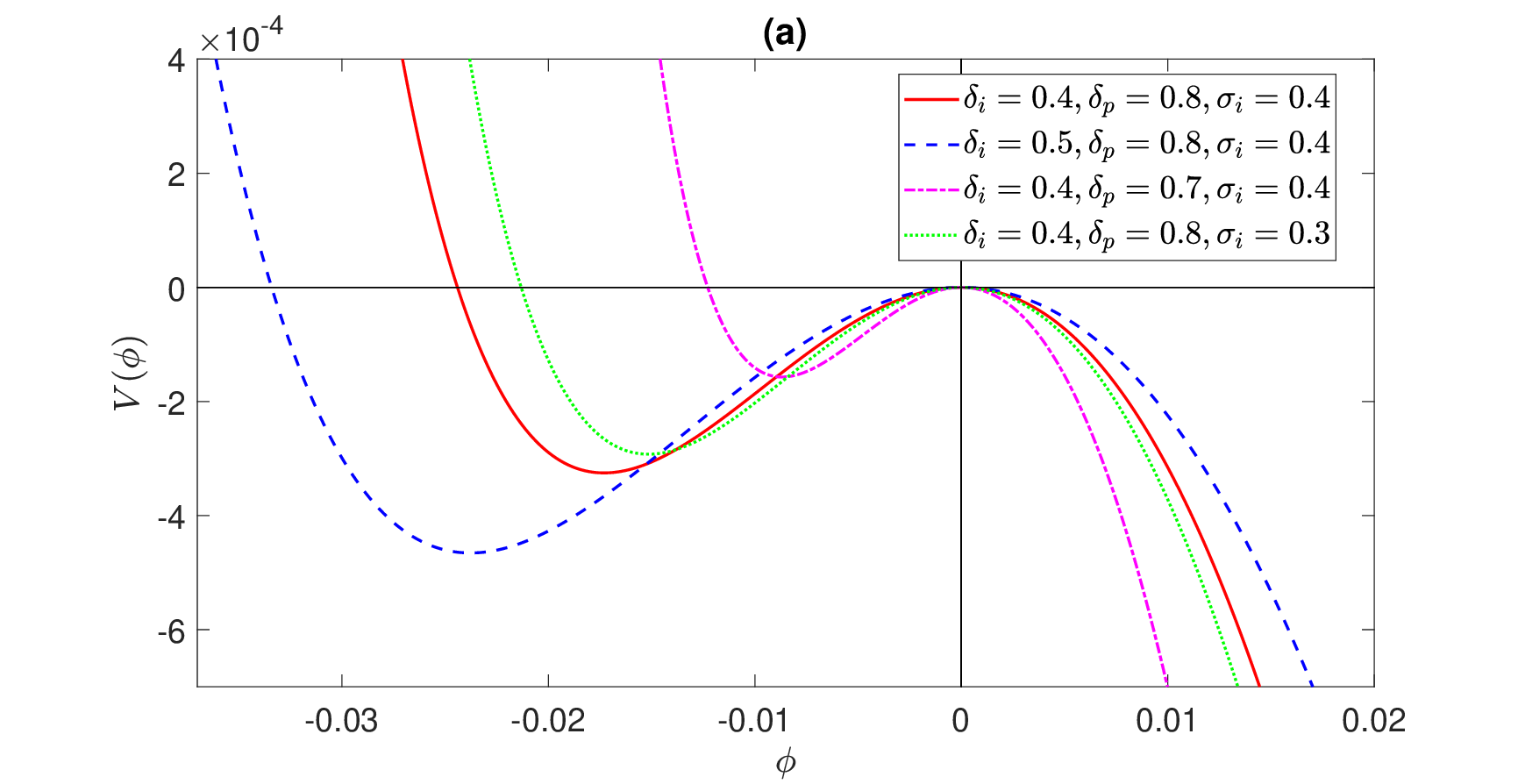}
\includegraphics[width=3.4in,height=2.5in]{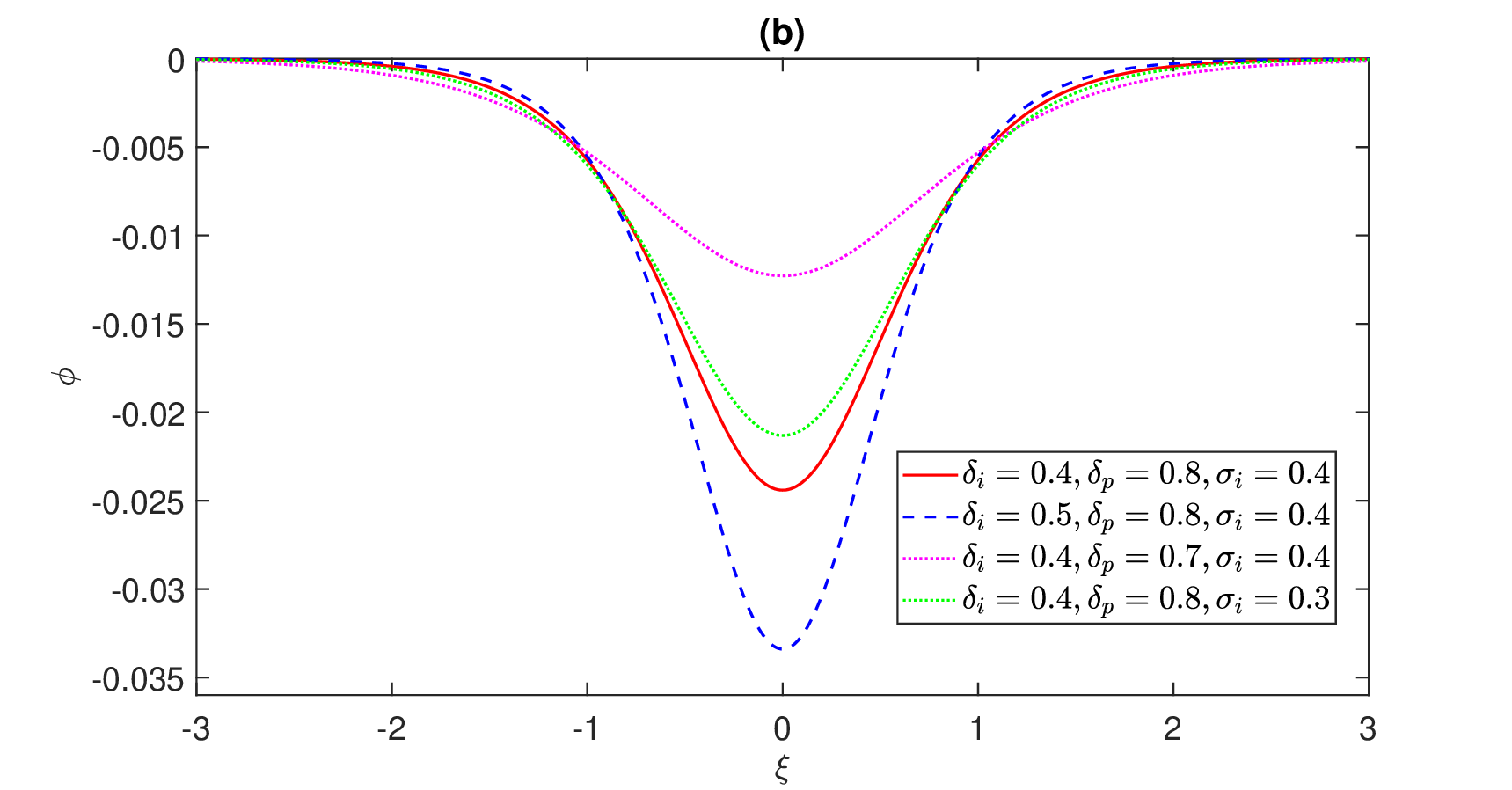}
\caption{(a) The Sagdeev pseudopotential $V(\phi)$ and (b) the corresponding solitary-wave potential profiles $\phi$ are shown for different values of $\sigma_i$, $\delta_i$ and $\delta_p$. The other parameters used are $\kappa_i=3$, $\tau_e=0.2$, $\sigma_p=1$ and $M/M_c=1.2$. }
\label{fig:Vphi_di_dp_si}
\end{figure}
\begin{figure}[!htbp]
\centering
\includegraphics[width=3.4in,height=2.5in]{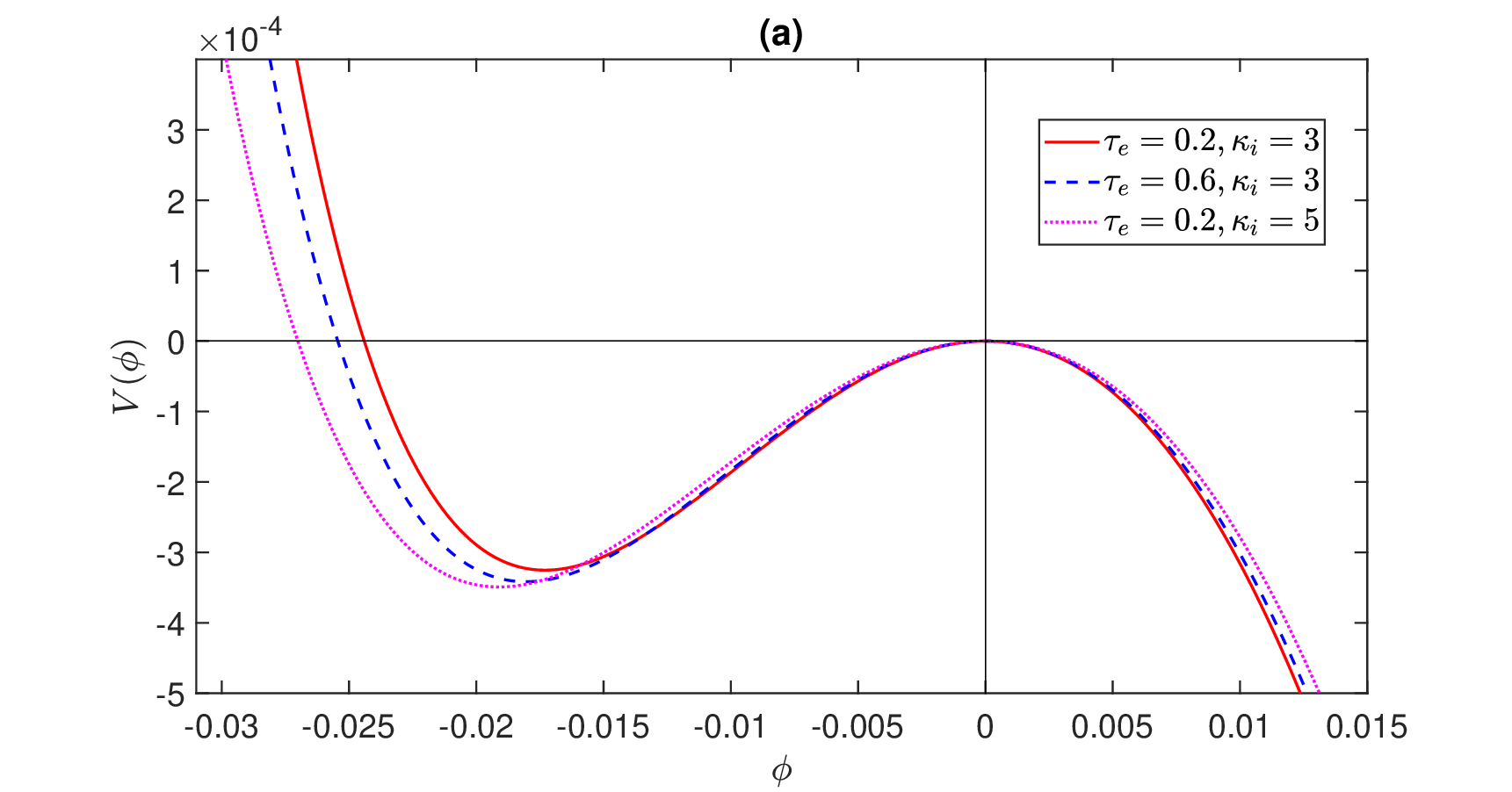}
\includegraphics[width=3.4in,height=2.5in]{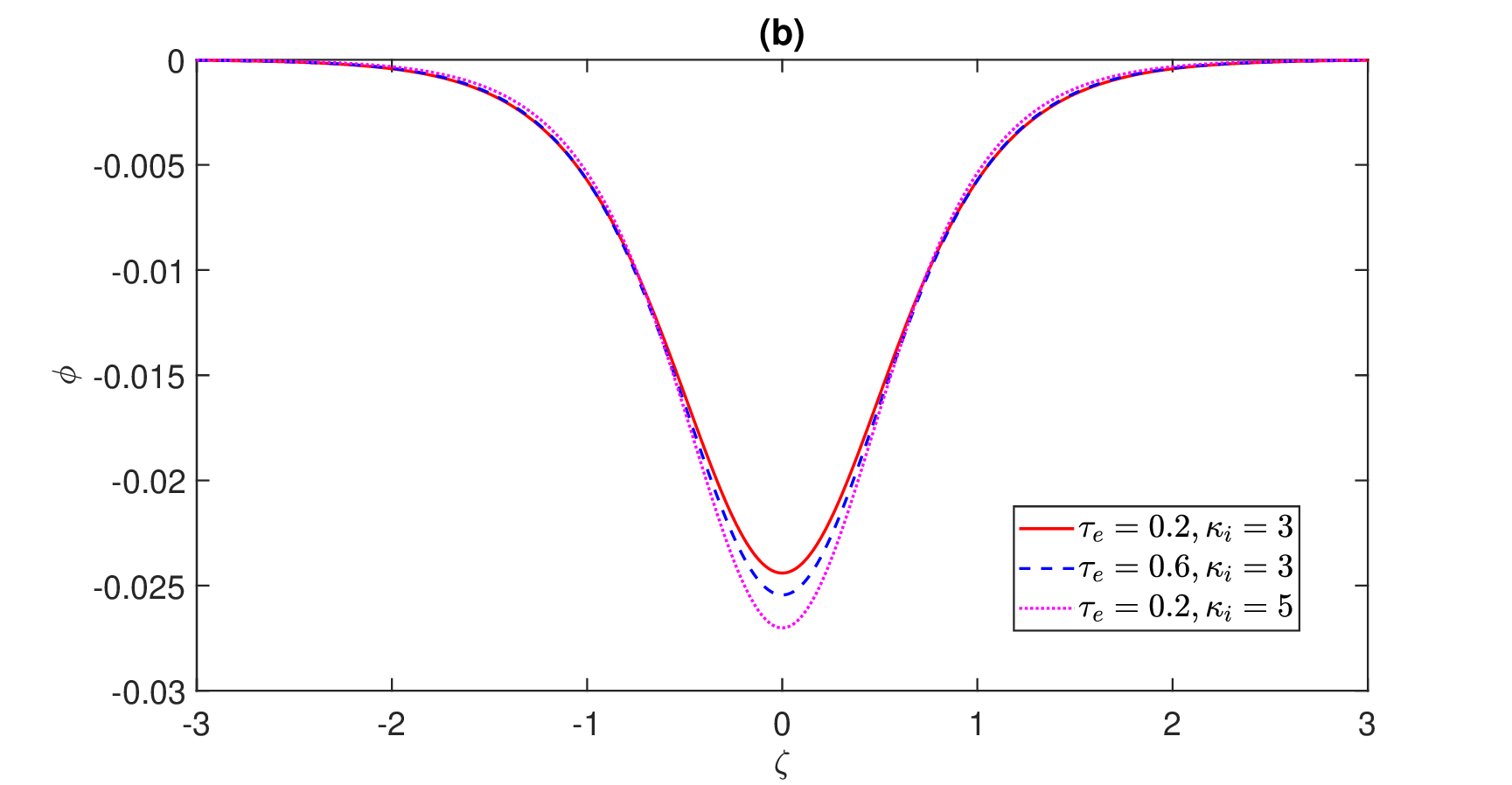}
\caption{ (a) The Sagdeev pseudopotential $V(\phi)$ and (b) the corresponding solitary-wave potential profiles $\phi$ are shown for different values of $\kappa_i$ and $\tau_e$. The other parameters used are $\sigma_i=0.4$, $\delta_i=0.4$ and $\delta_p=0.8$, $\sigma_p=1$ and $M/M_c=1.2$.}
\label{fig:Vphi_taue_ki}
\end{figure}
Having identified the parameter regimes for the existence of negative potential SWs, we now investigate their profiles, with particular emphasis on the dependence of amplitudes and widths on the plasma parameters. This requires an examination of both the Sagdeev potential profiles and the corresponding solitary wave solutions. The Sagdeev potential profiles $V(\phi)$s are obtained by plotting Eq. \eqref{eq-energy-integral} against the pseudoposition $\phi$, and the amplitude $\phi_m$ of the corresponding potential profiles can be identified. The width of the SW can be obtained either from the profile of $\phi(\zeta)$ obtained by integrating $d\zeta=d\phi/\sqrt{-2V(\phi)}$ or by computing the value of $\phi_m/{\sqrt{|V_m|}}$, where $|V_m|$ is the absolute minimum value of $V(\phi)$ in the pseudopotential wells. The pseudopotential $V(\phi)$ is presented in Fig. \ref{fig:Vphi_M}(a), and the corresponding potential profiles are shown in Fig. \ref{fig:Vphi_M}(b) for different values of the true Mach number $M/M_c$, while keeping all other parameters fixed as specified in the figure caption. All potential profiles in \ref{fig:Vphi_M}(b) exhibit localized bell-shaped depressions in negative potential solitary waves. The plots demonstrate that as the true Mach number increases beyond the critical value, the amplitude and width of solitary waves increase, exhibiting the characteristic behaviour of DA solitary waves in plasma systems.  Close to the critical Mach number ($M/M_c \to 1$), the solitary structures become weak and narrow, in agreement with the KdV-limit behaviour. This correspondence will be discussed in detail in the preceding section \ref{sec-small}. In Fig. \ref{fig:Vphi_di_dp_si}, $V(\phi)$ and the corresponding potential profiles are displayed for different values of $\delta_i$, $\delta_p$ and $\sigma_i$, while all other parameters are kept fixed as specified in the caption of the figure. The results demonstrate how variations in these plasma parameters influence the solitary waves and their amplitude–width characteristics. Fig. \ref{fig:Vphi_di_dp_si}(a) shows that the depth of the pseudopotential well and the corresponding wave amplitude are strongly influenced by variations in $\delta_i$, $\delta_p$ and $\sigma_i$. An increase in the ion density ratio $\delta_i$ enhances the effective charge imbalance, thereby strengthening the nonlinear response of the plasma and resulting in a deeper potential well. Similarly, a higher positron density ratio $\delta_p$ increases the restoring force of the plasma, thus supporting higher amplitude solitary waves. The increase in $\sigma_i$ also plays a significant role by increasing the thermal pressure of the ions, which modifies the balance between nonlinearity and dispersion, and ultimately contributes to the formation of sharper potential wells. These modifications in the Sagdeev potential are directly reflected in the solitary wave profiles of Fig. \ref{fig:Vphi_di_dp_si}(b). The profiles clearly reveal that as $\delta_i$, $\delta_p$ and $\sigma_i$ increase, the solitary waves acquire larger amplitudes while their spatial width decreases. To illustrate the influence of the temperature ratio $\tau_e$ and the ion superthermal index $\kappa_i$ on the characteristics of DASWs, in Fig. \ref{fig:Vphi_taue_ki}, the Sagdeev potential $V(\phi)$ (\ref{fig:Vphi_taue_ki}(a)) and the corresponding solitary wave profiles $\phi(\xi)$ (\ref{fig:Vphi_taue_ki}(b)) are plotted for different values of these parameters. From Fig. \ref{fig:Vphi_taue_ki}(a), it is evident that both $\tau_e$ and $\kappa_i$ substantially affect the depth of the Sagdeev potential. In particular, for a fixed $\kappa_i = 3$, increasing $\tau_e$ (solid red vs. blue dashed curve) leads to a deeper potential well and a shift in amplitude towards more negative values of $\phi$. This indicates that a higher temperature ratio allows the system to support solitary structures of larger amplitude. Similarly, for fixed $\tau_e = 0.2$, increasing $\kappa_i$ from $3$ to $5$ (red solid vs. magenta dotted curve) results in a higher amplitude. Since a higher $\kappa_i$ corresponds to a weaker deviation from the Maxwellian distribution, this suggests that strong superthermality (small $\kappa_i$) enhances the nonlinear trapping mechanism responsible for solitary wave formation. The corresponding solitary wave profiles in Fig. \ref{fig:Vphi_taue_ki}(b) are consistent with these observations. The results demonstrate that an increase in $\tau_e$ and $\kappa_i$ produces solitary waves with greater amplitudes and comparatively smaller widths.

\subsection{SMALL AMPLITUDE WAVE}\label{sec-small}
 \begin{figure}[!htbp] 
\centering
\includegraphics[width=3.8in,height=2.5in]{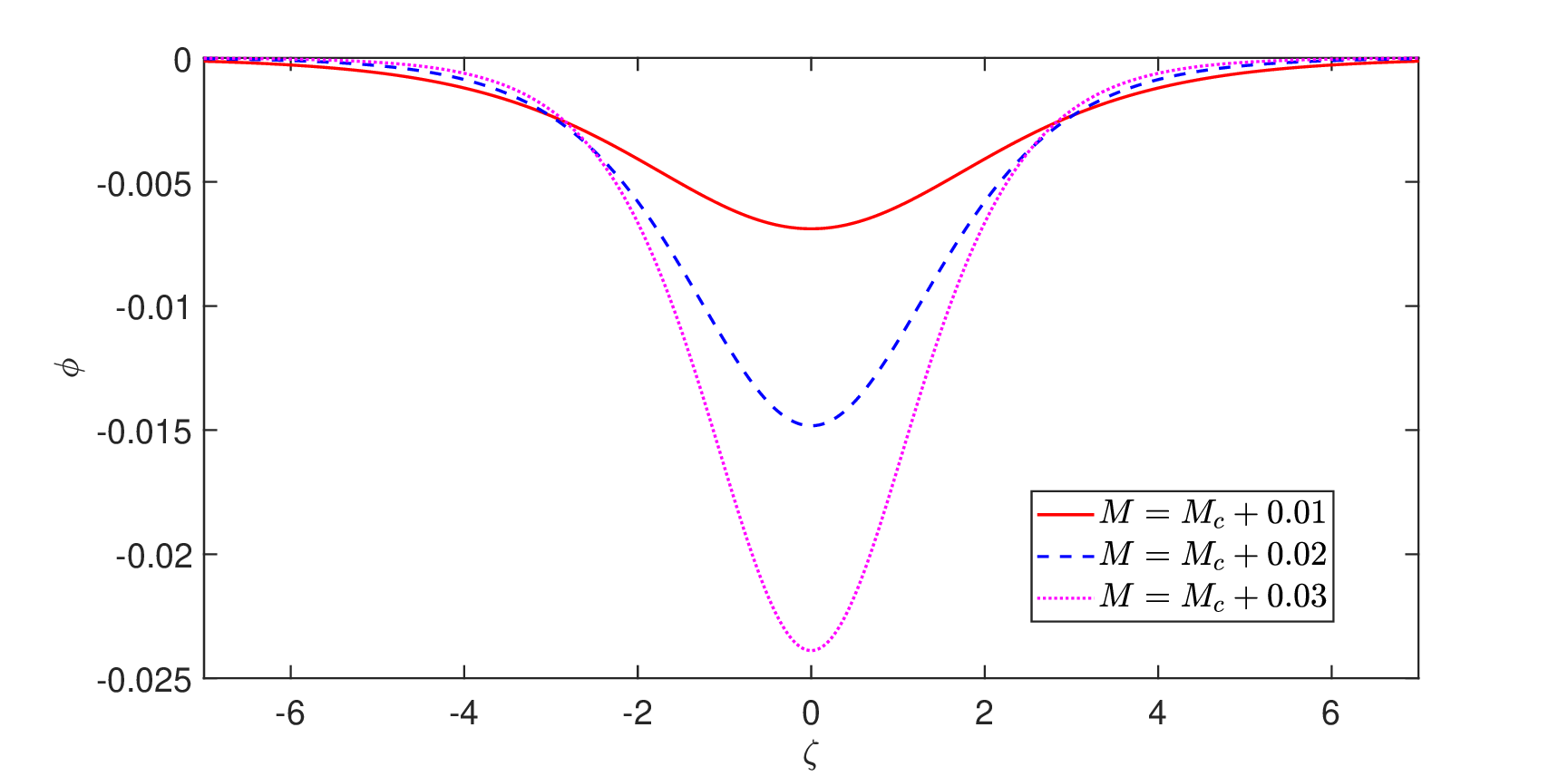}
\caption{The Profiles of small amplitude DASWs are shown for different values of mach number M. The other parameters used are $\sigma_i=0.4$, $\delta_i=0.4$ and $\delta_p=0.8$, $\sigma_p=1$, $\tau_e=0.2$ and $\kappa_i=5$.}.
\label{fig:small-soliton}
\end{figure}
Thus far, we have examined the existence domain and propagation characteristics of arbitrary-amplitude DASWs in terms of different plasma parameters relevant to dense plasma environments. While the arbitrary-amplitude formulation based on the Sagdeev potential provides a complete picture of the nonlinear structures, it does not explicitly yield analytical expressions for the soliton profiles. One particular point of interest in this context is to investigate the  SW profiles in the small-amplitude limit, where analytical approximations become tractable and physically insightful. In this regime, the pseudopotential $V(\phi)$ can be expanded in a Taylor series around $\phi = 0$ by assuming $\phi \ll 1$. Substituting this expansion into the energy integral \eqref{eq-energy-integral} and keeping upto third order terms of $\phi$, we obtain, 
\begin{equation} \label{eq-small-SW}
    \frac{1}{2}\left(\frac{d\phi}{d\zeta}\right)^2+ A{\phi^2}+B{\phi^3}=0,  
\end{equation}
  where the expansion coefficients $A$ and $B$ depend explicitly on the plasma parameters and are given by,
\begin{equation}\label{exp-A}
A= -\frac{\alpha_e}{2 \beta_e}-\frac{\alpha_p}{2\beta_p\sigma_p}-\frac{C_i\alpha_i}{2\sigma_i}+\frac{1}{2M^2}
\end{equation}
and
\begin{equation}\label{exp-B}
\begin{aligned}
B=-&\frac{\alpha_e}{6} \frac{Li_{-1/2}\left[-\exp(\mu_{e0})   \right]}{Li_{3/2}\left[-\exp(\mu_{e0})   \right]}\\&+\frac{\alpha_p}{6\sigma_p^2} \frac{Li_{-1/2}\left[-\exp(\mu_{p0})   \right]}{Li_{3/2}\left[-\exp(\mu_{p0})   \right]}\\ &+\frac{\alpha_iC_i}{6\sigma_i^2}\left(\frac{k_i+1/2}{k_i-3/2}\right)-\frac{1}{2M^4}.
\end{aligned}
\end{equation}
The resulting expression \eqref{eq-small-SW} usually corresponds to a Korteweg–de Vries (KdV) equation, where the amplitude and width of the solitary structure can be written in a closed analytical form. Furthermore, the width $w$ and amplitude $\phi_m$ of the soliton are connected by a relation $w^2\phi_m=2/B=constant$, which is a well-known relation derived from the KdV theory. Integrating Eq. \eqref{eq-small-SW} and using the boundary conditions $\phi \to 0$ and $ {d\phi}/{d\zeta} \to 0$ as $\zeta \to \pm\infty$, we obtain the usual DA soliton solution as follows
\begin{equation} \label{eq-soli-profile}
 \phi=\phi_m \mathrm{sech}^2\left({\zeta}/{w}\right),
\end{equation} 
where $\phi_m=-A/B$ and $w=(-2/A)^{1/2}$  are the DA soliton's amplitude and width, respectively. This approach is particularly important since it connects the fully nonlinear Sagdeev potential framework with the reductive perturbation method (RPM). Moreover, the small-amplitude analysis enables a clearer understanding of how various plasma parameters behave and influence the soliton amplitude and width, thereby providing a useful check against the arbitrary-amplitude results. For a real soliton solution \eqref{eq-soli-profile} to exist, one needs to have $A<0$ leading to the condition $M>M_c$ where $M_c$, given by \eqref{eq-lower-mach}, represents the linear phase velocity.
Since the small-amplitude approximation is valid only for a small deviation from the linear phase velocity, we assume $M=M_c+\varepsilon M_0$, where $\varepsilon$ is some small positive scaling parameter and $M_0$ represents the leading-order deviation from the phase velocity of the DA solitary waves. The pulse profile given by Eq. \eqref{eq-soli-profile} is identical to the solution of the Korteweg-de Vries (KdV) equation, which is obtained by using the RPM method in a small amplitude limit. The DA waves become non-dispersive and become a KdV-soliton at $M\approx M_c$. The nature of the DASW becomes rarefactive (or compressive) depending upon $B<0$ ($B>0$). $B>0$ leads to the condition $M>M_k$ where,
\begin{equation}\label{eq-M-k}
\begin{split}
      M_k&=\left[-\frac{\alpha_e}{3} \frac{Li_{-1/2} \left[-\exp\left(\mu_{e0}\right)\right]}{Li_{3/2} \left[-\exp\left(\mu_{e0}\right)\right]} \right.\\&\left. + \frac{\alpha_p}{3\sigma_p^2}\frac{Li_{-1/2} \left[-\exp\left(\mu_{p0}\right)\right]}{Li_{3/2} \left[-\exp\left(\mu_{p0}\right)\right]}\right.\\&\left.+\frac{\alpha_i C_i}{3\sigma_i^2}\left(\frac{k_i+1/2}{k_i-3/2}\right) \right]^{-1/4},
\end{split}
\end{equation}
implying $M_k-M_c<\varepsilon M_0$. However, the numerical investigations reveal that no physically admissible parameter region satisfies this condition. In view of the results presented in Section \ref{Sagdeev Potential}, where only rarefactive solitary structures of arbitrary amplitude were obtained, the possibility of $B>0$ may be excluded. A positive electrostatic potential would require excessive compression of the electrons and enhanced accumulation of negatively charged dust, both of which are strongly suppressed by degeneracy and thermal pressure effects in the present model. Since the negatively charged dust grains provide the inertia, while the electrons and positrons are inertialess and governed by strong degeneracy pressure, the Sagdeev pseudopotential develops a potential well only for negative potentials. Consequently, only rarefactive solitary waves are permitted in the system. For a choice of $M_0$ satisfying $\varepsilon M_0<M_k-M_c$, the propagation of small-amplitude DASW with negative potential becomes possible. The typical potential profiles obtained from \eqref{eq-soli-profile} are depicted in Fig.~\ref{fig:small-soliton} for Mach numbers taken slightly above the linear phase velocity $M_c$. The results confirm that as the Mach number increases marginally, the soliton amplitude increases while its width correspondingly decreases.  The qualitative characteristics of these solitons, with respect to variations in the system parameters, remain the same as those of arbitrary-amplitude waves. From a stability viewpoint, the existence of these localized structures is ensured by the presence of a well-defined Sagdeev pseudopotential well, within which the pseudoparticle executes bounded oscillatory motion about the unstable equilibrium point at $\phi=0$. Such bounded motion implies that the solitary structures are nonlinearly stable against small perturbations in the wave frame, as long as the Sagdeev potential retains a single well with $V(\phi)<0$ between $\phi=0$ and the maximum amplitude $\phi=\phi_m$. Since the present analysis admits well-defined Sagdeev potential wells only for negative potentials within the admissible Mach number range, the corresponding rarefactive DASWs are expected to be nonlinearly stable in the wave frame. A more detailed dynamical stability analysis based on time-dependent numerical simulations, however, is beyond the scope of the present work and is left for future studies.

\section{SUMMARY AND FUTURE OUTLOOK} \label{Conclusion}
In this work, we have investigated the propagation of DASWs in an unmagnetized, collisionless dusty e-p-i plasma where electrons and positrons are partially degenerate, ions follow a superthermal ($\kappa$) distribution, and cold negatively charged dust provides the inertia. Using linear analysis, we derived a generalized dispersion relation that incorporates the finite-temperature degeneracy of electrons and positrons through polylogarithm functions and the superthermal ion response through the $\kappa$ index. Nonlinear dynamics is treated with the Sagdeev pseudopotential method to obtain existence domains and arbitrary-amplitude solitary-wave solutions. A small-amplitude expansion is also presented to connect these results to the KdV (reductive perturbation) limit, which yielded analytical expressions for the soliton amplitude and width. The results obtained are validated against previous studies by reducing the present model to previously established limiting cases \cite{esfandyari2012large,saberian2017nonlinear}. The key physical mechanism underlying the formation of DASW in the present system is the balance between nonlinear dust inertia and electrostatic restoring forces modified by finite-temperature degeneracy of electrons and positrons and by ion superthermality. Partial degeneracy alters the pressure response of the light species through Fermi–Dirac statistics, thereby modifying the effective screening length and the critical Mach number, while superthermal ions introduce enhanced nonlinearity that reshapes the Sagdeev pseudopotential and constrains the admissible Mach-number domain. The combined action of these effects leads to the exclusive existence of rarefactive, subsonic dust–acoustic solitary structures with amplitudes and widths that are highly sensitive to the degeneracy strength, ion spectral index, and composition parameters. The main results can be summarized as
follows.

\begin{itemize}
    \item The linear phase speed and the dispersion behaviour are substantially modified by partial degeneracy of electrons and positrons and by the ion superthermality. Increasing electron degeneracy ($\tau_e$) or ion temperature ratio generally shifts the dispersion curves upward, while stronger superthermality (smaller $\kappa_i$) lowers the frequency for a given wave number.

    \item The Sagdeev analysis reveals that only negative-potential (rarefactive) DA solitary waves are supported in the parameter regimes studied. Solitary structures exist only for subsonic propagation relative to the DA speed in an admissible domain of the Mach number, bounded by the critical value $M_c$ and the upper limit $M_u$, i.e., the admissible true Mach number satisfies $1<M/M_c<M_u/M_c$. Although compressive solitary structures are not ruled out in general, the present Sagdeev analysis shows that, for the chosen plasma model and parameter ranges, only rarefactive solutions are supported.

    \item It is observed that the variation of plasma parameters changes both the amplitude and the width. The larger $\delta_i$, $\delta_p$, and larger $\tau_e$ or $\kappa_i$ deeper the Sagdeev well, producing higher-amplitude and narrower solitary waves. In contrast, stronger superthermality (smaller $\kappa_i$) and cooler ions widen the existence domain but tend to reduce the amplitude for a fixed true Mach number.
\end{itemize}
The results have direct relevance for dense astrophysical and space environments—such as white dwarf envelopes, neutron star magnetospheres, and dusty accretion regions where partially degenerate electron–positron populations can coexist with superthermal ions and charged dust  \cite{dolai2020effects, akbari2011nonlinear, farihi2016circumstellar, zobaer2012nonlinear}. The analysis highlights how finite-temperature degeneracy and nonthermal ion tails modify collective electrostatic structures in such systems and can alter observationally relevant properties, viz., speed, amplitude and width of DA pulses. In summary, this study demonstrates that the partial degeneracy of electron-positron and ion superthermality plays a significant and nontrivial role in the shaping of DASWs: they determine the existence domain, fix subsonic propagation limits, and control the properties of amplitude and width. The combined Sagdeev and small-amplitude analyses provide a coherent picture that can serve as a baseline for more realistic and extended models of nonlinear electrostatic structures in dense dusty plasmas. While plasma behavior outside the considered parameter ranges, such as full degeneracy at extremely high densities or alternative distribution functions, may exhibit different nonlinear features.

 Furthermore, the present analysis is restricted to an unmagnetized and collisionless regime, which is appropriate for a broad class of dusty plasma environments. The effects of magnetic fields and collisions will be addressed in future studies. A magnetic field would primarily introduce anisotropy in the dust dynamics, modify the dispersion characteristics through cyclotron effects, and alter the Sagdeev pseudopotential structure via an effective magnetic pressure term. Also, extending the present model to include external forces, such as gravity, will be considered in future work. The inclusion of finite dust temperature would introduce an additional dust pressure contribution, and a systematic investigation of warm-dust effects and their quantitative impact on the existence domain and soliton profiles remains another direction for research. Effects associated with dust charge fluctuations and finite dust temperature are neglected in this study and will be addressed in future studies. Although the present work is focused on one-dimensional DASWs within the Sagdeev pseudopotential framework, the underlying balance between dispersion and nonlinearity considered here is closely related to the mechanisms responsible for lump-wave formation in higher dimensions. Therefore, the analytical approach adopted in this study may be extendable to multi-dimensional nonlinear plasma models and integrable systems, where rational and lump-type solutions naturally arise. Time-dependent numerical simulations of the full fluid–Poisson system could offer additional support by illustrating the dynamical formation and stability of the predicted solitary waves in the present model. Incorporating numerical simulations to complement the Sagdeev analysis will be addressed in future studies.

\begin{table}[t] 
\caption{NOMENCLATURE}
\begin{tabular}{lll} 
\hline 
Symbol/ \\ Abbreviation & Description & Unit\\ 
\hline 
$j=e, p, i, d$ & Electron, positron, ion & -- \\ 
 & and dust, respectively& \\ 
$n_j$ & Number density of the $j$th species & cm$^{-3}$ \\ 
$n_{j0}$ & Unperturbed number density & cm$^{-3}$ \\ 
 & of the $j$th species &  \\ 

$u_d$ & Dust particle velocity & cm s$^{-1}$\\ 
$T_j$ & Thermodynamic temperature & K \\
& of the $j$th species &  \\ 
$E_{Fe} (E_{Fp})$ & Fermi energy of electron (positron) & erg \\ 
$k_B$ & Boltzmann constant& erg K$^{-1}$\\ 
$T_{Fe} (T_{Fp})$ & Fermi temperature & K \\ 
& of electron (positron) &  \\
$\mu_{e0}(\mu_{p0})$ & Unperturbed chemical potential   & erg  \\
  & of electron (positron) &   \\
$\sigma_p(\sigma_i)$ & Temperature ratio $T_p/{T_e} (T_i/{T_e})$ &--\\ 
$\phi$ & Electrostatic potential & StatV\\
$\hbar$ &Planck's constant& erg s\\
$\tau_e (\tau_p)$ & Degeneracy parameter & --\\
& of electron (positron) & \\
$\kappa (\kappa_i)$ & Superthermality index (for ions)& --\\
$c_d$& Modified dust acoustic speed& cm s$^{-1}$\\
$Z_d$&The number of charged dust particles&--\\
$p_e(p_p)$ & Electron (Positron) fluid pressure& dyne cm$^{-2}$\\
$\omega_{pd}$ & Dust plasma frequency &  s$^{-1}$ \\ 
$\lambda_{D}$ & Debye length & cm \\ 
$\delta_p(\delta_i)$ & Density ratio $n_{p0}/{n_{e0}} (n_{i0}/{n_{e0}})$&--\\ 
$\omega$ & Angular frequency & s$^{-1}$ \\  
$k$ & Wave number& cm$^{-1}$\\
$M$ & Mach number normalized by $c_d$ & -- \\
$M_c (M_u)$ & Critical (Upper limit) Mach number& -- \\ 
DA & Dust-acoustic & -- \\ 
DAWs & Dust-acoustic waves & -- \\ 
DASW & Dust-acoustic solitary wave & -- \\ 
DASWs & Dust-acoustic solitary waves & -- \\
SW & Solitary wave &--\\
SWs & Solitary waves &--\\  
FD & Fermi-Dirac & -- \\
e-p & Electron-positron & -- \\ 
e-p-i & Electron-positron-ion & -- \\  
KdV & Korteweg-de Vries &--\\ 
RPM & Reductive perturbation method &--\\
\hline 
\end{tabular} 
\end{table}

\section*{ACKNOWLEDGEMENTS}
  The authors gratefully acknowledge the anonymous referees for their insightful and constructive comments that significantly improved the quality of this work.

\section*{DATA AVAILABILITY}
Data sharing is not applicable to this article as no new data were
created or analyzed in this study.

\bibliographystyle{apsrev4-2}
\bibliography{References}

\end{document}